\documentstyle[preprint,aps,epsf]{revtex}
\tightenlines
\begin{document}
\draft
\title{\bf From bound states to resonances: \\ 
analytic continuation of the wave function} 
\author{G. Cattapan and E. Maglione}
\address{Istituto Nazionale di Fisica Nucleare e 
Dipartimento di Fisica dell' Universit\'a\\
via Marzolo 8, Padova I-35131, Italia}

\maketitle

\begin{abstract}
Single--particle resonance parameters and wave functions in spherical
and deformed nuclei are determined through analytic continuation in
the potential strength. In this method, the analyticity of the eigenvalues 
and eigenfunctions of the Schr\"odinger equation with respect to the
coupling strength is exploited to analytically continue the bound--state
solutions into the positive--energy region by means of Pad\'e approximants
of the second kind. The method is here applied to single--particle wave
functions 
of the $^{154}{\rm Sm}$ and $^{131}{\rm Eu}$ nuclei. A comparison of the
results with the direct solution of the Schr\"odinger equation shows that  
the method can be confidently applied also in coupled--channel situations
requiring high numerical accuracy.
\end{abstract}


\narrowtext

In recent years there has been an increasing experimental activity\cite{wd97}
on nuclei far from the stability line. These nuclear systems
are usually unbound or weakly bound, and often exhibit resonances
with a pronounced single--particle character, since the Fermi level is
close to or even immersed in the continuum. This phenomenology has
prompted people to look for more and more efficient methods for the
description of unbound states, which could compete with the
techniques presently available for the discrete part of nuclear
spectra. A possible approach, which proved to be rather successful, 
exploits the fact that resonances can be described by wave functions 
with purely outgoing behavior and complex eigenvalues $E_R - i\Gamma/2$ 
(Gamow states) \cite{vl95}. One is then confronted with the solution of a 
complex eigenvalue problem for a non--Hermitian Hamiltonian. Even if 
Gamow states are characterized by the at first sight unpleasant feature 
of an exponentially growing oscillatory behavior at large distances, 
they can be however normalized by a suitable generalization of the 
quantum--mechanical inner product. This can be achieved in several, 
essentially equivalent ways, either introducing convergence factors in 
the integrals expressing their norm \cite{ze61,gv71}, or by analytic 
continuation from the upper half of the $k$--plane to the resonance poles 
\cite{ho65,ro68}. These recipes can be implemented by solving the
Schr\"odinger equation along a deformed contour in the complex $r$--plane
\cite{gv71}, which amounts to an analytic continuation ${\bf r} \rightarrow 
exp{(i\theta)}{\bf r}$ in configuration space. This is very convenient
under a computational point of view, resonance states being transformed 
into square--integrable wave functions, while leaving untouched 
the corresponding poles of the S--matrix in the energy plane. If this
is done for a value of $r$ so large that the effective nucleon--core
interaction can be neglected (exterior complex scaling), one has the 
advantage that the potential itself remains unchanged in the contour 
deformation. Complex--coordinate rotation methods have found interesting 
applications both in nuclear \cite{vc89} and in atomic \cite{ho83} physics.  

Recently, unbound states in exotic nuclei have been studied by means of 
Analytic Continuation in the Coupling Constant (ACCC) \cite{ts97,ts99}. This 
method, proposed already several years ago by Kukulin and co--workers 
\cite{kk77,kk79}, starts from the intuitive expectation that, for an 
attractive potential, a resonance state will become a bound state as the 
coupling strength is increased. Under a mathematical point of view, one 
can prove that the wave number $k = \sqrt{2 \mu E}/\hbar$ is an analytic 
function of the strength $\lambda$, with the same restrictions on the 
potential which guarantee the analyticity of the Jost function 
\cite{kk79,ta72}. Here and in the following $\mu$ denotes the reduced mass
of the nucleon--core system. Near the value $\lambda_0$ for which 
$k(\lambda_0) = 0$ ({\it i.e.} near the scattering threshold), one has 
\cite{kk79,ta72}
\begin{equation}
k(\lambda) \sim i\sqrt{\lambda - \lambda_0} \quad , \quad 
k(\lambda) \sim i(\lambda - \lambda_0) , 
\label{kll}
\end{equation}
for $l \neq 0$, and for $l = 0$, respectively. These properties suggest the 
analytic continuation of $k$ in the
complex $\lambda$--plane from the bound--state region into the resonance
region through the employment of Pad\'e approximants of the second kind
\cite{kk79}
\begin{equation}
k \simeq k^{(N,M)}(x) = i {c_0 + c_1 x + c_2 x^2 + \cdots + c_M x^M
\over {1 + d_1 x + d_2 x^2 + \cdots + d_N x^N}} ,
\label{pad}
\end{equation}   
where $x \equiv \sqrt{\lambda - \lambda_0}$. In practice, the following
procedure can be followed to find the resonance parameters 
for an interaction $V$, when $l \neq 0$. One endows $V$ 
with a strength parameter $\lambda$, $V \rightarrow \lambda V$, and 
solves the bound--state problem for $\lambda V$ in correspondence to 
$N+M+1$ different values $\lambda_i$ of the coupling strength. Given 
the threshold value $\lambda_0$, the $N+M+1$ coefficients in the Pad\'e 
approximant (\ref{pad}) can be determined by equating $k^{(N,M)}(x_i)$
to the actual values $k_i$ of the wave number. The approximant can then 
be used to estimate the resonance wave number $k_{r}$, and hence the
resonance position and width, in correspondence to the ``physical" value
$\lambda =1$ of the potential strength. 
This procedure can be easily modified if the potential $V$ supports
an $s$--wave resonance, in which case the corresponding pole leaves
the negative imaginary axis at $k(\bar \lambda_0) = -i\bar \chi_0$
\cite{ts97,ts99,kk79}. The method has been applied with some success 
to unbound states in $^5$He and $^5$Li, as well as in three--cluster 
nuclei \cite{ts97,ts99}. An extrapolation procedure similar the one described 
above can be used in the complex $k$--plane to analytically continue 
the bound--state wave function $\psi_l^{(B)}(kr)$ into the scattering 
region for any value of the radial variable $r$.

        In this paper we study the application of this technique
to single--particle resonances in deformed nuclei, a situation
quite common in the drip--line region. The numerical evaluation 
of resonance (Gamow) states for nonspherical nuclei is a major challenge, 
which has been only recently solved \cite{fm97}. Bound--state wave
functions, on the other hand, can be calculated nowadays through
very efficient and quick algorithms \cite{cd87}. Here, we shall enquire
whether the solution of the coupled--channel bound--state
problem, obtained for a set of values of the coupling strength
$\lambda$, can be analytically continued into the unbound region.
We shall consider the Pad\'e extrapolation both for the resonance 
parameters and for the wave function. The latter case is particularly
interesting for proton decay, since one has to deal with lifetimes
$\tau \geq 1 \mu s$, which implies resonance widths $\Gamma =
\hbar/\tau$ smaller than $10^{-16}$ MeV. Such small widths are
difficult to obtain with enough accuracy starting directly from the 
energy eigenvalue, and can be best estimated from the wave--function
behavior \cite{ma98,ma99}. 

        To test the validity of the ACCC method in the present case, 
we have solved exactly the problem both for bound--state energies and 
in the continuum, and we have compared the results with the outcome 
of the Pad\'e extrapolation obtained starting from the bound--state 
region, for different, decreasing values of the coupling strength. The
radial Schr\"odinger equation with outgoing--wave boundary conditions
has been solved in the standard way, namely starting from the origin 
and from the outer region, and matching the logarithmic derivative of 
these functions at some radius $R$. For real, negative energies one
gets normalized wave functions, with the proper exponentially
decreasing tail, in correspondence to the energy eigenvalues of the bound 
system; in the scattering case, on the other hand, one gets purely
outgoing states for complex eigenvalues $E_R - i\Gamma/2$ \cite{vl95}. The
normalization of our unbound wave functions agrees with the Zel'dovich
\cite{ze61} or Gyarmati--Vertse \cite{gv71} prescriptions for Gamow states.

As a first step, let us consider single--particle resonances in spherical 
nuclei. In Fig.\ \ref{fig1} we report the results of our calculations for
the $f_{5/2}$ neutron state in the nucleus $^{154}{\rm Sm}$. The neutron
is assumed to move in a spherical Saxon--Woods potential, supporting a
bound state for $\lambda =1$. Fig.\ \ref{fig1}(a) exhibits the real  
and imaginary part of the energy $E$, as a function of the
decreasing strength parameter $\lambda$. The full dots represent the 
outcome of the numerical solution of the Schr\"odinger equation, whereas 
the crosses correspond to the bound--state energies used for the
Pad\'e extrapolation into the scattering region, the extrapolated
results being given by the full lines. As it can be seen, the 
neutron is bound for $\lambda$ decreasing from 1 down to $\lambda_0 \simeq 
0.84$, where the neutron state becomes unbound and the energy acquires a 
non--vanishing imaginary part. With a (7,7) approximant of the form 
(\ref{pad}), the agreement between the exact and the extrapolated results 
is quite good both for $Re(E)$ and $Im(E)$ in the whole considered
region of the strength--parameter values. In Figs.\ \ref{fig1}(b) and 
\ref{fig1}(c) we show the calculated (full dots) and extrapolated values of 
the real and imaginary part of $r\psi_{lj}(r)$ at $r = 7$ fm and $r = 15$ 
fm, respectively. The results are plotted as functions of the real part 
$Re(E)$ of the corresponding eigenvalue. As in Fig.\ \ref{fig1}(a), the 
crosses denote the bound--state points used as input to the extrapolation 
procedure. Note that at $r \sim 7$ fm the neutron is feeling the strongest
effect from the nuclear potential, and the wave function is attaining its 
largest value; for $r \sim 15$ fm, on the other hand, one is far away enough
from the nuclear core, to have an indication of the quality of the 
extrapolation in reproducing the tail of the wave function. This is 
crucial in order to obtain an accurate evaluation of the resonance width.
Indeed, in the spherical case the partial width for the decay to the
channel $lj$ can be related to the value of the wave function 
$\psi_{lj}(r)$ at $r =R$ by \cite{ma98}
\begin{equation}
\Gamma_{lj} = {\hbar^2 k \over \mu} {R^2 |\psi_{lj}(R)|^2 \over
{|G_{l}(R) + i F_{l}(R)|^2}} ,
\label{wid}
\end{equation}
where $F_{l}$ and $G_{l}$ represent the free regular and irregular
radial wave functions, respectively. If $R$ is large enough to be 
outside of the range of the potential, Eq. \ref{wid} provides 
a width which is actually independent from the value of $R$, as 
it should be \cite{ma98}.  These considerations can be extended to the 
coupled--channel case \cite{ma99}. As Figs.\ \ref{fig1}(b) and \ref{fig1}(c)
clearly exhibit, the results of the (7,7) extrapolation agree very well 
with the numerical calculation both in the inner and in the tail region.
In particular, the non--trivial behavior of the real part of the inner 
wave function in the threshold region is accurately reproduced by the 
extrapolation, an indication that the Pad\'e analytic continuation performs
well in transferring the information from the bound--state into the scattering
region. Similarly, the steep increase of the wave--function tail when going
to positive energies is successfully reproduced by the Pad\'e approximation.
We have also considered the behavior of the Pad\'e extrapolation for
lower--rank approximants, and verified that the analytic continuation still
compares well with the exact results when a (4,4) Pad\'e approximant is
employed.

	Similar calculations have been done for a proton state in the
$^{154}{\rm Sm}$ nucleus, with the proton moving in a $h_{9/2}$ state. In
this case, when solving the Schr\"odinger equation, the inner wave 
function has to be matched to the radial Coulomb wave function 
$r\psi_{lj}^{out}(r) = N_{lj} [G^{(C)}_{l}(r) + i F^{(C)}_{l}(r)]$, where 
$G^{(C)}_{l}$ and $F^{(C)}_{l}$ are the usual regular and irregular Coulomb 
functions and $N_{lj}$ is a normalization factor. The outcome of the 
calculations is reported in Fig.\ \ref{fig2}. As for the neutron case, 
Fig.\ \ref{fig2}(a) displays the real part of the energy eigenvalue for
decreasing $\lambda$, whereas Figs.\ \ref{fig2}(b) and \ref{fig2}(c) refer
to the real part of $r\psi_{lj}(r)$ at $r = 5$ fm and $r = 15$ fm, 
respectively, the symbols having the same meaning as in Fig.\ \ref{fig1}.
The imaginary parts of both the energy and the wave function remain 
vanishingly small in passing from the bound--state into the scattering 
region, so that they are not even reported in the figures. The proton state
becomes unbound at $\lambda \sim 1$, with $Re(E)$ scaling linearly with the
strength parameter. Similarly, the wave function exhibits a smooth behavior
when the scattering threshold is crossed. It is not surprising, therefore,
that a low--rank (4,4) Pad\'e approximant can provide an extremely good
reproduction of the exact results. Note that, because of the Coulomb barrier,
the proton wave function in the exterior region is much smaller than in the
neutron case.

	The ACCC method has been finally tested for proton resonances
in deformed nuclei. As is well--known, such a calculation is much more
challenging than in the spherical case, since it requires the solution
of a coupled--channel problem to determine the intrinsic, single--particle
wave functions \cite{fm97,ma98,ma99}. In Fig.\ \ref{fig3} (left panel) 
we report the real part of the energy of the $3/2+$ state in $^{131}{\rm Eu}$
as a function of $\lambda$. Improved experimental data have been recently
given for this highly deformed proton emitter, leading to the identification
of a fine structure splitting in the radioactive decay from the ground
state \cite{sd99}. The process can be successfully described in the
framework of a model where the proton is emitted from a deformed 
single--particle Nilsson level \cite{ma99,mf99}. The energy of the
$3/2+$ state turns out to be a linear function of $\lambda$, much as in 
the spherical situation, the imaginary part remaining always very close to 
zero. The relevant wave--function components at $r= 15$ fm are given in 
the right panel of Fig.\ \ref{fig3}. For the proton decay of the 
$^{131}{\rm Eu}$ nucleus from the $K = 3/2+$ state they are the $d3/2$ and
$d5/2$ components of the $3/2+$ Nilsson orbital (higher--$l$ partial waves
giving a negligible contribution to the decay width \cite{mf99}). The former
determines the $^{131}{\rm Eu}$ proton decay to the $^{130}{\rm Sm}$ ground
state, while the latter is by far the dominant term in the decay to the 
$2^+$ excited state of the daughter nucleus \cite{ma99,mf99}. The bound--state
solution has been analytically continued into the scattering region by means
of a simple (3,3) Pad\'e approximant. One can see that the extrapolation 
reproduces the outcome of the numerical solution extremely well both for the
energy and for the wave--function components. Note that, in passing from the
bound--state into the positive--energy region the imaginary part of the wave
functions is again so small, that it is not given in the figure.

	In summary, we have applied a Pad\'e extrapolation to determine 
the resonance parameters and wave function for single--particle resonances
in nuclei, starting from bound--state calculations. This has been obtained
by varying the coupling strength so as to analytically continue the
bound--state energy eigenvalue and wave function into the 
positive--energy region. The method has been applied to the single--particle
decay of both spherical and deformed nuclei, which entails the solution 
of a single-- or a coupled--channel problem, respectively. A comparison 
of the extrapolated results with the outcome of the direct solution of 
the Schr\"odinger equation with the proper boundary conditions shows that 
the ACCC method can be confidently applied to these situations, where
high numerical accuracy is required in order to have a meaningful
comparison with the experimental data.

\begin{figure}
\caption{Energy eigenvalue and wave function  for the neutron $f5/2$ state 
in the $^{154}{\rm Sm}$ nucleus. In (b) and (c) the wave function 
has been multiplied by the radius $r$, and evaluated at $r = 7$ fm and 
$r = 15$ fm, respectively. Full dots represent the results obtained from
the numerical solution of the Schr\"odinger equation, whereas crosses are
the input values used for the analytic continuation. Full curves are the
outcome of the Pad\'e extrapolation.}
\label{fig1}
\end{figure}

\begin{figure}
\caption{As in Fig.\ \ref{fig1} for the real part of the energy and
of the quantity $r\psi_{lj}(r)$ referring to the proton $h9/2$ state
in the $^{154}{\rm Sm}$ nucleus. In (b) and (c) the wave function is 
evaluated at $r = 5$ fm and $r = 15$ fm, respectively.}
\label{fig2}
\end{figure}

\begin{figure}
\caption{As in Fig.\ \ref{fig2} for the proton $3/2+$ state in the
deformed nucleus $^{131}{\rm Eu}$. The dominant wave--function
components $r\psi_{lj}(r)$ are evaluated at $r = 15$ fm.}
\label{fig3}
\end{figure}

\newpage

\epsfysize= 18 truecm 
\epsffile[50 50 700 700]{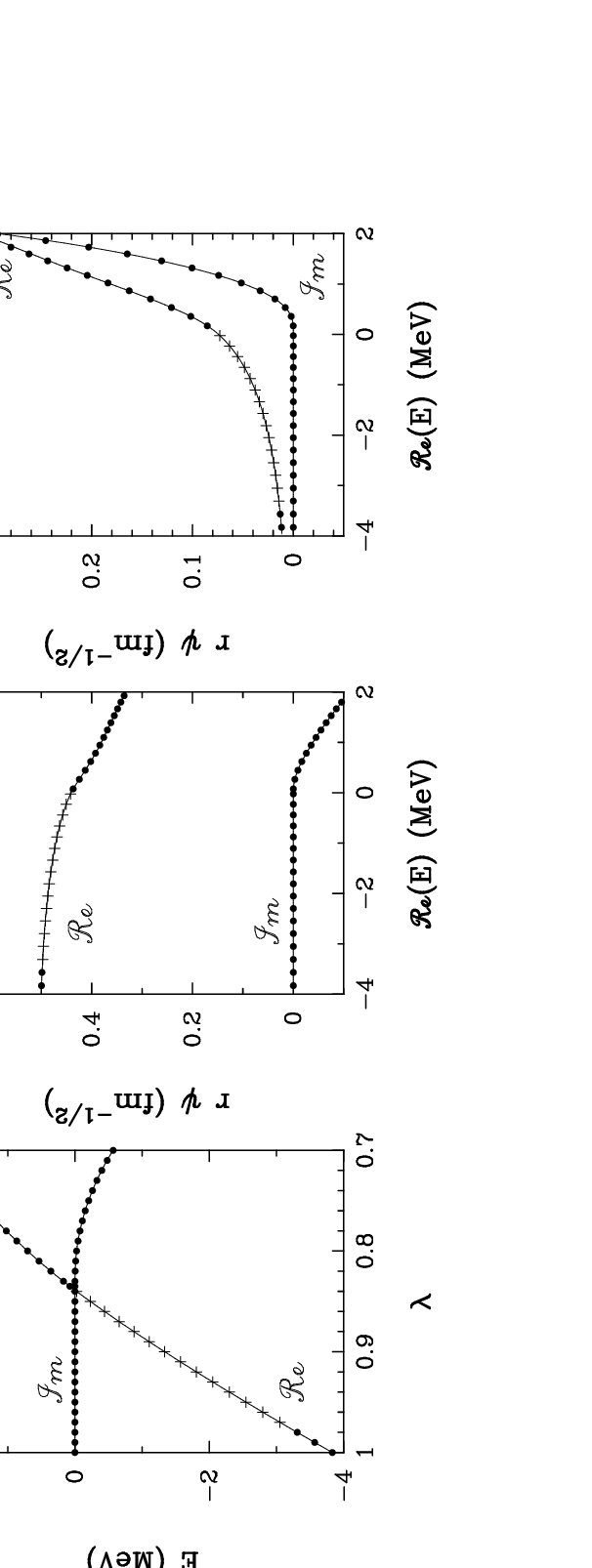} 

\newpage

\epsfysize= 18 truecm
\epsffile[50 50 700 700]{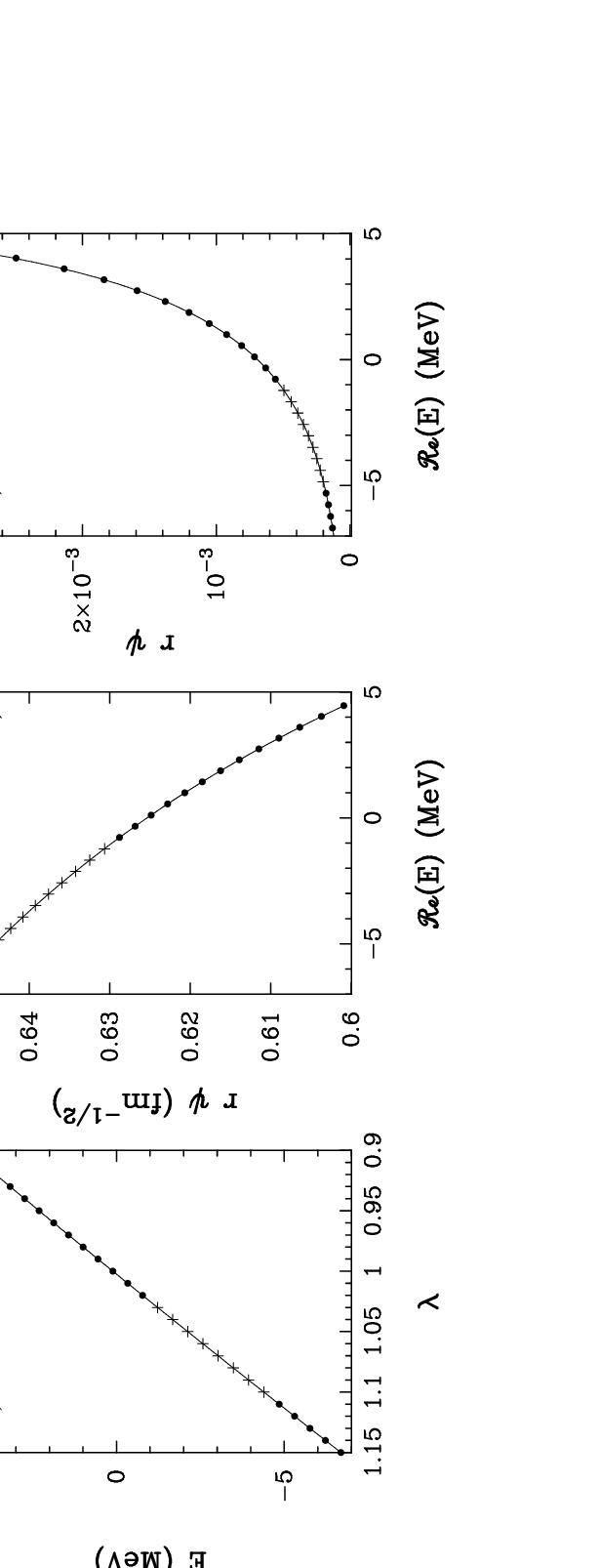}

\newpage

\epsfysize= 18 truecm
\epsffile[50 50 700 700]{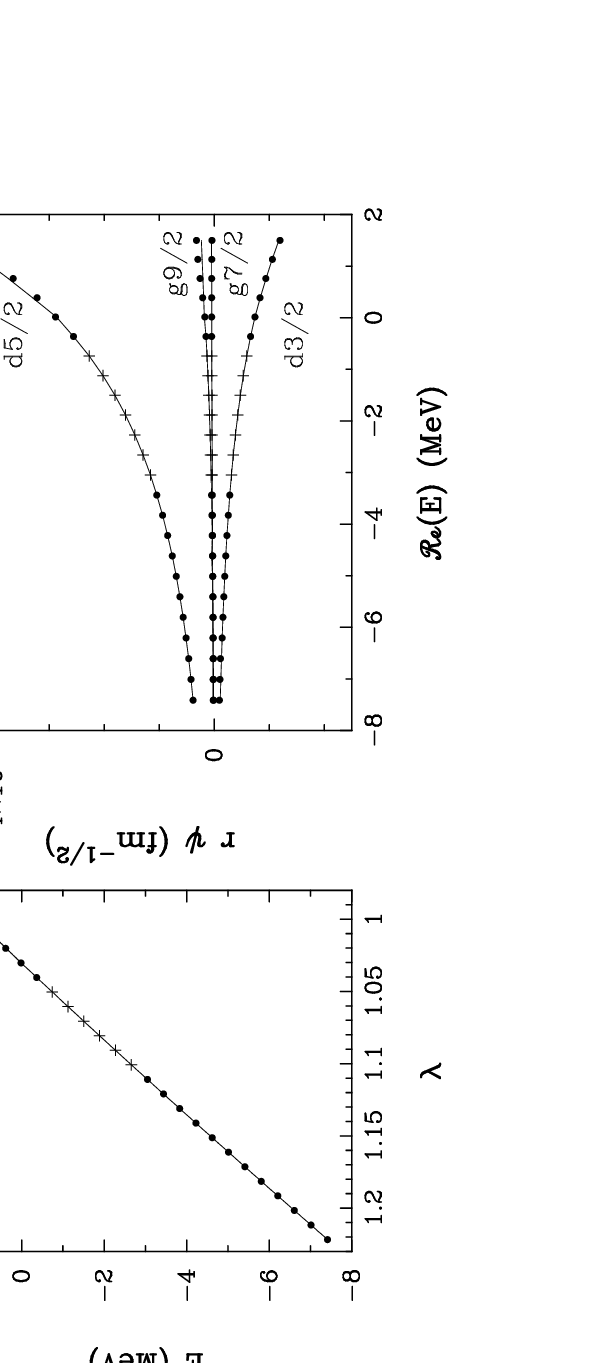}

\end{document}